\documentclass{elsarticle}

\usepackage{amssymb}
\usepackage{amsmath}
\usepackage{program}
\usepackage{multirow}
\usepackage{rotating}
\usepackage{url}
\usepackage{color}

\newcommand{\geneig}{GSYEIG}
\newcommand{\stdeig}{STDEIG}

\newcommand{\nref}[1]{(\ref{#1})}
\renewcommand{\R}{\mathbb{R}}
\newcommand{\Rnn}{\R^{n \times n}}
\newcommand{\Rss}{\R^{s \times s}}
\newcommand{\Rns}{\R^{n \times s}}

\newcommand{\ph}[2]{\stackrel{\phantom{#1}}{#2}}

\begin{document}

\begin{frontmatter} 

\title{Solving 
  Dense 
  Generalized Eigenproblems \\
  on Multi-threaded Architectures
}
            
\author[uji]{%
Jos\'e I. Aliaga%
}

\author[aachen]{%
Paolo Bientinesi%
}

\author[ruder]{%
Davor Davidovi\'c%
}

\author[jsc]{%
Edoardo Di Napoli%
}

\author[uji]{%
Francisco D. Igual\corref{cor1}%
}

\author[uji]{%
Enrique~S.~Quintana-Ort\'{\i}%
}

\address[uji]{%
 Depto. de Ingenier\'{\i}a y Ciencia de Computadores,
    Universidad Jaume I,
    12.071--Castell\'on, Spain.
    {{aliaga,figual,quintana}@icc.uji.es}
}

\address[aachen]{%
 RWTH-Aachen University,
    52056--Aachen, Germany.
    {{pauldj@aices.rwth-aachen.de}}
}

\address[ruder]{%
 Institut Ruder Bo\u{s}kovi\'c,
    Centar za Informatiku i Ra\u{c}unarstvo - CIR,
    10000--Zagreb, Croatia.
    {{ddavid@irb.hr}}
}

\address[jsc]{%
 JSC,
     Forschungszentrum J\"ulich,
     52275--J\"ulich, Germany.
     {{dinapoli@aices.rwth-aachen.de}}
}

\cortext[cor1]{Corresponding author}

            

\begin{abstract} 
We compare two approaches to compute a fraction of the spectrum of 
dense symmetric definite generalized eigenproblems: one is based on the
reduction to tridiagonal form, and the other on the Krylov-subspace
iteration.  Two large-scale applications, arising in molecular dynamics and
material science, are employed to investigate the contributions of the
application, architecture, and parallelism of the method to the
performance of the solvers. 
The experimental results on a state-of-the-art 
8-core platform, equipped with a graphics processing unit (GPU), reveal
that in realistic applications, iterative Krylov-subspace methods can be a 
competitive approach also for the solution of dense problems.
\end{abstract}


\end{frontmatter} 

\section{Introduction}

We consider the solution of the {\em generalized eigenproblem}
\begin{equation}
\label{eqn:gsyeig}
A X = BX\Lambda,
\end{equation}
where $A,B \in \Rnn$ are given, $\Lambda \in \Rss$ is a diagonal matrix with the 
$s$ sought-after eigenvalues, and the columns of $X \in \Rns$ contain
the corresponding unknown eigenvectors~\cite{GVL:1996}. 
When the pair ($A, B$) consists of a symmetric and a  
symmetric positive definite matrix, Eq.~\eqref{eqn:gsyeig} is normally 
referred to as a {\em symmetric-definite generalized eigenproblem} (\geneig).
We are interested in large-scale \geneig s 
arising in the simulation of molecular dynamics~\cite{Chacon2011} 
and {\em ab initio} simulations of materials~\cite{FLEUR}; 
in these applications, $A$ and $B$ are symmetric and dense,
$B$ is positive definite (SPD), 
$n \approx {\cal O}(10,000)-{\cal O}(100,000)$, 
and only few eigenpairs (eigenvalues and associated eigenvectors) 
are required: $s \ll n$.

For the solution of \geneig s with dense $(A,B)$, 
there exist two numerically stable approaches: 
the ``{\em tridiagonal-reduction}'' and 
the ``{\em Krylov-subspace iteration}''~\cite{GVL:1996}.
Both of them start by transforming ---either explicitly or implicitly--- 
the generalized problem~\nref{eqn:gsyeig} into a {\em standard} one (\stdeig).
Specifically, consider the Cholesky factorization of $B$ given by 
\begin{equation} 
\label{eqn:cholesky} 
B=U^TU, 
\end{equation} 
where $U \in \Rnn$ is upper triangular~\cite{GVL:1996}; 
then the \geneig\ can be transformed into the \stdeig 
\begin{equation} 
\label{eqn:syeig}
   C Y = Y\Lambda,
\quad \equiv \quad
   (U^{-T} A U^{-1}) (U X)  = (UX) \Lambda, 
\end{equation}
where $C \in \Rnn$ is symmetric, and $Y \in \Rns$ contains the eigenvectors associated
with this problem. While the eigenvalues of the \geneig~\nref{eqn:gsyeig} and the 
\stdeig~\nref{eqn:syeig} are the same, the eigenvectors $X$ of \geneig\ can be
easily recovered from those of \stdeig, $Y$, 
by solving the upper triangular linear system 
\begin{equation}
\label{eqn:backone}
X := U^{-1} Y.
\end{equation}

After this preliminary transformation, 
the tridiagonal-reduction approach employs orthogonal transforms 
to reduce~$C$ to tridiagonal form, from which the eigenpairs 
can be computed.
On the other hand, the Krylov-subspace approach operates with $C$ 
(either directly or via the matrices $A$ and $U$), iteratively 
approximating the largest (or smallest) eigenpairs of the system
through matrix-vector multiplications.
When dealing with dense coefficient matrices,
both families of numerical methods exhibit a computational cost of 
${\cal O}(n^3)$ floating-point arithmetic operations (flops), 
due to the transformation to standard form and, in the tridiagonal-reduction 
approach, the reduction to tridiagonal form.
Therefore, the solution of large-scale dense eigenproblems, 
as those appearing in molecular dynamics
or {\em ab initio} simulations, clearly
calls for the application of high-performance 
computing techniques on parallel architectures.

Traditionally, the tridiagonal-reduction approach has been regarded as
the method-of-choice for the solution of dense eigenvalue problems
while the Krylov-subspace alternative was preferred for sparse
matrices. However, as we will show in this paper, on parallel architectures,
the Krylov-subspace method is a competitive option for the solution of dense eigenvalue
problems, and the adoption of one method over the other should be
based instead on a variety of factors, such as the number of
required eigenpairs and the target architecture.

The major contribution of this paper is an experimental study of these
two classes of numerical eigensolvers, implemented using parallel
linear algebra libraries and kernels for current desktop
platforms, for two large-scale applications.  Following the
evolution of computer hardware, we include two distinct architectures
in the evaluation: A system equipped with (general-purpose)
multi-core processors from Intel, and a hybrid computer that embeds
multi-core processors with (one or more) NVIDIA ``Fermi'' GPUs
(graphics processor units).  For brevity, we will refer to both
multi-core processors and GPUs as multi-threaded architectures.  The
linear algebra libraries include well-known packages like
LAPACK~\cite{lapack} or BLAS, as well as alternatives for
multi-threaded architectures like PLASMA, {\tt libflame} or
MAGMA~\cite{plasmaweb,flameweb,magmaweb}.


The rest of the paper is structured as follows.  In
Section~\ref{sec:solvers} we review the different eigensolvers that
are considered in this work, offering a brief description of the
underlying numerical methods and their computational and storage
costs.  In Section~\ref{sec:setup} we describe the experimental setup:
The two large-scale applications leading to dense \geneig s, and the hybrid
multi-core/GPU platform on which we conduct the experiments. In
Section~\ref{sec:conventional_libraries} we revisit the numerical
methods, now from the point of view of conventional software libraries
(LAPACK, BLAS, SBR, ARPACK) that can be employed to implement them,
and evaluate these implementations on the target multi-core processor,
via the two case studies.  We then repeat the experimentation using
more recent libraries, specifically designed to leverage
task-parallelism and/or hardware accelerators like the GPUs in
Section~\ref{sec:modern_libraries}.  A short discussion of concluding
remarks as well as future work is provided in
Section~\ref{sec:conclusions}.

\section{Generalized Symmetric Definite Eigenvalue Solvers\label{sec:solvers}}

In this section we first review the initial transformation from 
\geneig\ to \stdeig, and the final back-transform. 
We then describe the two approaches for the solution of \stdeig, 
---tridiagonal-reduction and Krylov-subspace iteration---
and a number of algorithmic variants.
We will assume that, initially, the storage available to the methods consists of 
two $n \times n$ arrays  (for the data matrices $A$ and $B$), 
and an $n \times s$ array (for the requested $s$ eigenvectors). 
Hereafter we neglect the space required to store the $s$ sought-after 
eigenvalues as well as any other lower order terms in storage costs. 
Analogously, in the following we neglect the lower order terms 
in the expressions for computational costs.

\subsection{Transformation to and from \stdeig}

The initial factorization in~\nref{eqn:cholesky} requires $n^3/3$
flops, independently of $s$, the number of eigenpairs requested.  In
practice, the triangular factor $U$ overwrites the corresponding
entries in the upper triangular part of $B$ so that the demand for
storage space does not increase.  The algorithmic variants of the
tridiagonal-reduction approach require the matrix $C:= U^{-T}AU^{-1}$
to be explicitely built; the same is true for one of the variants of
the Krylov-subspace approach. 
In all cases, the entries of $A$ can be overwritten with the result $C$.
The computational cost for this operation amounts to
$2n^3$ flops if $C$ is computed by solving two triangular linear systems.
By exploiting the symmetry of $C$, the cost can instead be reduced to
$n^3$ flops; again, the cost is independent of $s$. Conversely, the
final back-transform~\nref{eqn:backone} costs $n^2s$ flops, and this
operation can be performed in-place.

\newcommand{\trid}{{\sc td}}
\newcommand{\trit}{{\sc tt}}
\newcommand{\krye}{{\sc ke}}
\newcommand{\kryi}{{\sc ki}}

\subsection{Tridiagonal-reduction approach}

Once \geneig\ has been transformed to \stdeig, 
we consider two alternative methods for reducing $C$ to tridiagonal form.
The first one performs the reduction in a single step, while the
second employs two (or possibly more) steps, reducing the full and dense $C$  
to a banded matrix, and from there to the required tridiagonal form.

\paragraph{Variant \trid:} {\em \underline{T}ridiagonal-reduction with 
\underline{D}irect tridiagonalization.}
Efficient algorithms for the solution of Eq.~\nref{eqn:syeig} usually
consist of three stages. Matrix $C$ is first reduced to
symmetric tridiagonal form by a sequence of orthogonal similarity
transforms: $Q^T C Q = T$, where 
$Q \in \Rnn$ is the matrix obtained from the accumulation of the orthogonal transforms,
and $T \in \Rnn$ is the resulting tridiagonal
matrix. 
In the second stage, a tridiagonal eigensolver, for instance the MR$^3$
algorithm~\cite{Dhillon20041,bientinesi:dhillon:rvdg}, is
employed to accurately compute the desired $s$ eigenvalues of 
$T$ and the associated eigenvectors.
In the third and last stage, a back-transform yields the eigenvectors of $C$; 
specifically, if $T Z = Z \Lambda$, with $Z \in \R^{n \times s}$
containing the eigenvectors of $T$,
then $Y := Q Z$. 
The first and last stages cost $4n^3/3$ and $2n^2s$ flops,
respectively. The complexity of the second stage, when performed by the MR$^3$ algorithm,
is ${\cal O}(n s)$ flops for computing $s$ eigenpairs.
(Other alternatives for solving symmetric tridiagonal eigenproblems, 
such as the QR algorithm, the Divide \& Conquer method,
etc.~\cite{GVL:1996} require ${\cal O}(n^3)$ flops in the worst case,
and are rarely competitive with the MR$^3$ algorithm~\cite{PetPB2011}.)

In this three-stage method, the orthogonal matrix $Q$ is never constructed;
the corresponding information is implicitly stored in the form of
Householder reflectors in the annihilated entries of $C$. Therefore,
for this first variant of the tridiagonal-reduction approach,
there is no significant increase of the memory demand.

\paragraph{Variant \trit:} {\em \underline{T}ridiagonal-reduction with 
\underline{T}wo-stage tridiagonalization.}
One major problem of variant \trid\ is that half
of the computations required to reduce $C$ to tridiagonal
form are performed via Level 2 BLAS operations; these
operations are considerably less efficient than the Level 3 BLAS kernels,
especially on current multi-threaded architectures.

An alternative to obviate this problem
is to perform the reduction in two steps, 
first transforming the matrix $C$ from dense to band form 
($W \in \Rnn$, with bandwidth $w$) 
and then from band to tridiagonal form.
Provided $(32 \leq)\ w \ll n$, this allows casting most 
computations during the reduction process
($Q_1^TCQ_1 = W$, $Q_2^TWQ_2 = T$)
in terms of efficient Level 3 BLAS operations, 
at the expense of a higher computational cost.
(The choice of the value 32 is based on 
experimental experience with this algorithm~\cite{CPE:CPE1680}. In particular, increasing this blocking factor permits
a better reuse of cached data; however, it rapidly raises the computational cost of the 
subsequent reduction from band to tridiagonal form. Therefore,
a balance between these two factors is needed.)
In particular, reducing the matrix to tridiagonal by such a two-stage method basically
requires $4n^3/3$ flops to obtain $W$, and a lower-order amount for refining that into $T$. 
However, due to this double-step, recovering $Y$ from the eigenvectors of $T$, as
$Y := Q_1 Q_2 Z$, adds $7 n^3/3 + 2 n^2 s$ flops to the method, and thus yields a much higher 
cost than for the previous alternative. Specifically,
the full $n \times n$ matrix $Q_1$ needs to be explicitly constructed, 
which requires $4 n^3/3$ flops. Then,
one needs to accumulate $Q_1Q_2$, for $n^3$ flops, and finally calculate $(Q_1 Q_2)Z$, 
for an additional $2n^2s$ flops. 

In practice, the accumulation of $Q_1$ is done by multiplying the identity matrix from the right with a sequence of the orthogonal
transforms that are required to reduce panels (column blocks) of the input matrix to banded form.
Therefore, these accumulations can be completely performed via Level 3 BLAS operations (explicitly, via two calls to the matrix-matrix
product per panel as the orthogonal transforms are applied by means of the WY representation).
On the other hand, during the reduction from band to tridiagonal form, the matrix $Q_2$ is not explicitly constructed, but accumulated from the right
into the previously constructed $Q_1$. Although the reduction itself does not proceed by blocks, the accumulation of the orthogonal transforms
are delayed to introduce blocked operations for the update.
Therefore, the construction of $Q_1Q_2$ is fully cast in terms of Level 3 BLAS operations.

The banded matrix $W$ can be saved in compact form overwriting $n \times w$ entries of~$A$.
Unfortunately, in this approach we need to explicitly build the full matrix $Q_1$, which requires space
for an additional $n \times n$ array. 

\subsection{Krylov-subspace approach}
\label{subsec:krylov}

Instead of reducing matrix $C$ to tridiagonal form, one can employ (a variant of)
the Lanczos procedure~\cite{GVL:1996} to iteratively construct an orthogonal basis of the Krylov subspace associated
with $C$. 
At each iteration, by using a recursive three term relation, the procedure calculates a
tridiagonal matrix $T_m$ of dimension $m \times m$, with $2s\leq m \ll n$, whose
extremal eigenvalues approximate those of $C$, and a matrix $V_m$ of dimension $n
\times m$ with the corresponding Krylov vectors. If the eigenvalues of $T_m$ accurately approximate 
the $s$ sought-after eigenvalues of $C$, the iteration is stopped. Otherwise, the best $s$ approximations
are used to restart the Lanczos procedure~\cite{arpackweb}.
Under certain conditions,  and especially for symmetric matrices, 
the process often exhibits fast convergence. 

Despite the simplicity of the Lanczos procedure, due to floating point arithmetic,
 the orthogonality between the column vectors of $V_m$ is rapidly lost.
As a consequence, once an eigenvalue has been found, the algorithm might fail 
to ``remember'' it, thus creating multiple copies. 
A simple 
method to overcome this issue is to perform the orthogonalization 
twice, as suggested by Kahan in his unpublished work and later demonstrated by formal analysis~\cite{Giraud}.
Alternatively, orthogonality can be monitored during the construction of the subspace, ``ammending'' it in case it is lost.
Re-orthogonalizing Lanczos vectors once adds a variable computational cost to the algorithm, which can be up to $O(mn)$ in the worst scenario.
The cost of obtaining the eigenpairs from $T_m,V_m$  is $O(m^2)$ flops.
Moreover, the dimension of the auxiliary storage space required in these methods is of the order of $n \times m$ or smaller.

\paragraph{Variant \krye:} {\em \underline{K}rylov-subspace with \underline{E}xplicit 
construction of $C$.} 
Like in the two variants of the previous approach, variant
\krye\ explicitly builds the matrix $C$, as illustrated in
Eqn~\eqref{eqn:syeig}.
Each iteration $k$ of the Krylov subspace method then performs a (symmetric)
matrix-vector product of the form
$z_{k+1}:=Cw_k$, with $z_{k=1,2,\ldots} \in \R^n, {w}_{k=0,1,\ldots} \in \R^n$, and $w_0$ an initial guess,
requiring $2n^2$ flops per product. While obtaining
$w_{k+1}$ from $z_{k+1}$ only requires a few operations of linear cost in $n$, 
the re-orthogonalization (in case it is needed)
has an cost that varies between $O(n)$ and $O(mn)$ flops (best and worst case) 
and, potentially, can contribute substantially to the total computational time already for moderate values of $m$. 
In addition, a (data-dependent) number of implicit restarts are needed after
the Lanczos augmentation step, each involving the application of the QR iteration to the tridiagonal matrix $T_m$, 
thus resulting in a cost of $O(nm^2)$ flops per restart. 

\paragraph{Variant \kryi:} {\em \underline{K}rylov-subspace with \underline{I}mplicit operation on $C$.} 
In this variant, the matrix $C$ is not formed. Instead, 
at each iteration of the iterative method, 
the calculation $z_{k+1}:=U^{-T}AU^{-1}w_k$ is
performed 
as a triangular system solve, followed by a matrix-vector
product and, finally, a second triangular system solve: 
$z_{k+1}:=U^{-T}(A(U^{-1}w_k))$.  In this variant, there is no initial
cost to pay for the explicit construction of $C$, but the cost per
iteration for the computation of $z_{k+1}$ doubles with respect to the
previous case, from $2n^2$ to $4n^2$ flops.
In the iteration, obtaining $w_{k+1}$ from $z_{k+1}$ requires $O(n)$ flops, 
and the aforementioned re-orthogonalization costs $O(nm)$ flops;
in addition, each of the restarting steps performs $O(nm^2)$ flops.

\section{Experimental Setup\label{sec:setup}}

In this section we briefly introduce the two benchmark applications that require the solution
of 
dense \geneig\
and the platform on which we carried out the numerical experiments.

\subsection{Molecular Dynamics}

The first application-generated \geneig\ appears in molecular simulations of biological systems using normal mode analysis (NMA) in internal coordinates. 
Normal mode analysis (NMA) merged with coarse-grained models (CG) has proven to be a powerful and popular alternative of standard molecular dynamics to 
simulate large collective motions of macromolecular complexes at extended time scales~\cite{Cui05,Tam06,Skj09}.  
In the approach, biomolecule atomic degrees of freedom are treated explicitly in solving the generalized eigenvalue problem in a biologically relevant conformation. 
The computed eigenvalues, also known as modes, form an orthonormal basis of displacements, i.e. any biomolecule conformational change can be expressed as a 
linear combination of the modes.  Furthermore, excellent correlation has been found between the motion characterized only by the low frequency modes and the 
experimentally observed functional motions of large macromolecules. In the approach leading to the data matrices for this example, a recent implementation 
delivers the NMA low frequency modes by using 
dihedral angles as variables and employing different multi-scale CG representations~\cite{Chacon2011}. This very efficient tool has been applied successfully 
to predict large-scale motions enzymes, viruses, and large protein assemblies from a single conformation~\cite{Chacon2011}.  
As illustrative case, in this case we used the biological 
relevant low frequency modes using default parameters. This system comprises $n$=9,997 internal coordinates to be solved in a 
generalized  eigenproblem with both $A$ and $B$ SPD matrices. 
For the characterization of the collective motion, only about 1\% of the smallest eigenpairs are needed. 
In order to accelerate the convergence of the Lanczos iteration of the Krylov-subspace approach, in the experiments we compute the largest eigenpairs of the 
inverse problem $B X = AX\Lambda^{-1}$.

\subsection{Density Functional Theory simulations}


The second eigenproblem appears within an {\em ab initio} simulation arising in
Density Functional Theory (DFT), one of the most effective frameworks for studying
 complex quantum mechanical systems at the core of materials science.
  DFT provides the means to solve a
  high-dimensional quantum mechanical problem by transforming it into a large set of coupled one-dimensional equations,
  which is ultimately represented as a non-linear generalized eigenvalue problem.
  The later is solved self-consistently through a series of successive
  iteration cycles: the solution computed at the end of one cycle is used to generate the input in the next
  until the distance between two successive solutions is negligible.

  Typically a simulations requires tens of cycles before reaching convergence.
  After the discretization -- intended in the general sense of reducing a continuous
  problem to one with a finite number of unknowns -- each cycle comprises dozens
  of large and dense \geneig s $P^{(i)}_{\bf k}: A^{(i)}_{\bf k} x - \lambda B^{(i)}_{\bf k} x$
  where $A$ is Hermitian and $B$ Hermitian positive definite.
  Within every cycle,
  the eigenproblems are parametrized by the reciprocal lattice vector ${\bf k}$,
  while the index $i$ denotes the iteration cycle.
  The size of each problem ranges from 10,000 to 40,000 and the interest lies in the
  eigenpairs corresponding to the lower 10-20\% part of the
  spectrum; the solution of such eigenproblems is one of the most time-consuming stages in the entire simulation.

The problem solved in this paper comes from the simulation of
the multi-layer material GeSb$_2$Te$_4$, one of the phase-changing materials
used in rewritable optical discs (CDs, DVDs, Blu-Rays discs)
and prototype non-volatile memories.
The matrices (carrying indices $i=10$ and ${\bf k}=1$) were originated with the FLEUR code~\cite{FLEUR}
at the Supercomputing Center of the Forschungszentrum J\"ulich.
The size of the eigenproblem is
$n=$17,243 and the number of eigenpairs searched for is
$s=$448, corresponding to the lowest 2.6\% of the spectrum.

\subsection{Target platform}

The experiments were carried out using 
double-precision arithmetic on a platform 
equipped with two Intel Xeon Quadcore processors E5520 (8 cores at 2.27 GHz), 
with 24 GBytes of memory, connected to an NVIDIA  Tesla C2050 (Fermi) 
GPU (480 cores at 1.15 GHz) with 3 GBytes of on-device memory. 
The operating system is the $64$-bit CentOS $5.4$.
The following software libraries were employed:
ARPACK 1.4.1, 
CUBLAS 4.0, 
CUDA driver 4.0, 
Intel MKL 10.3, 
GotoBLAS2 1.11, 
{\tt libflame} 5.0, 
MAGMA 1.0 RC5, 
PLASMA 2.4.2, 
and SBR 1.4.1.
Codes were compiled using \texttt{gcc} 4.1.2 and/or \texttt{gfortran} 4.1.2 with the \texttt{-O3} optimization flag.

A large effort was made to optimize parameters 
like the block size of the various routines,
the bandwidth for variant \trit, the number of Krylov vectors ($m$) for \krye\ and \kryi, etc.
For the Krylov-subspace methods, the stopping threshold of routine {\sc dsaupd} was set to
the default ({\tt tol}=0). Internally, the code accepts the 
computed eigenpairs if the estimated relative residuals are below the machine precision.

\section{Conventional Libraries for Multi-core Processors\label{sec:conventional_libraries}}

\subsection{Exploiting multi-threaded implementations of BLAS}

In the dense linear algebra domain, the traditional approach to exploit the concurrency of 
a platform equipped with multiple processors (or cores) relies on the usage of 
highly-tuned, multi-threaded implementations of BLAS, often provided by the hardware vendors
(Intel's MKL, AMD's ACML, IBM's ESSL, etc.)
or by independent developers (e.g., GotoBLAS2). 
During the past decade, this was successfully leveraged by 
LAPACK~\cite{lapack} as well as {\tt libflame}~\cite{libflameref}
to yield acceptable speed-ups with no effort on the programmer's side. 

The combination of LAPACK and BLAS provides most of the
functionalities required to construct all the four algorithms (\trid, \trit, \krye, \kryi)  
to solve \geneig s on multi-core processors; see
Table~\ref{tab:keys}. Although LAPACK provides a specific routine to
construct $C:=U^{-T}AU^{-1}$ ({\sc dsygst}), in our tests we found
that computing $C$ via two triangular system solves ({\sc dtrsm}) was
faster; therefore this is the option selected in our
implementations. 
 
The missing components are provided by the SBR (Successive Band
Reduction) toolbox~\cite{Bischof:2000:AST} and the ARPACK
library~\cite{arpackweb}.  The former contains software for reducing a
full/band matrix to band/tridiagonal form via orthogonal similarity
transformations (variant \trit), while the later implements an
implicitly restarted version of the Lanczos iteration (to obtain
$w_{k+1}$ from $z_{k+1}$; see variants \krye\ and \kryi\ in
subsection~\ref{subsec:krylov}).  In the SBR toolbox, parallelism can
be obtained using a multi-threaded BLAS. Conversely, the benefits of
a parallel execution of ARPACK ---which mostly performs Level 1 and 2
BLAS operations--- will not be as significant.
In principle, the computation of the eigenvalues of the tridiagonal matrix 
$T_m$ and the eigenvectors from the Krylov vectors in $V_m$
add a minor cost to the overall computation due to the reduced value of $m$
compared with the dimension of the problem.
ARPACK employs a modified version of the symmetric iterative QR algorithm
for this purpose~\cite{GVL:1996}.



\newcommand{\supermatrix}{{\tt lf+SM}}

\newcommand{\num}[2]{{{\sc #1}}{\footnotesize{#2}}}

\begin{table}
{\footnotesize 
\begin{center}
\begin{tabular}{|c|c|c|cc|c|c|} 
\hline
Stage & Appr.  &  Var. &      & Operation & Routine & Library\\ \hline \hline
\multirow{2}{*}{{\scriptsize (1)}}  & \multirow{2}{*}{--}       & \multirow{2}{*}{--} & \num{gs}{1}
			& $\ph{.}{B}=U^TU \rightarrow U$ & {\sc dpotrf} & LAPACK\\ 
                      &            & & \num{gs}{2}     & $C:=U^{-T}AU^{-1}$ & {\sc dsygst}/{\sc dtrsm} & {LAPACK/BLAS} \\ \hline
\multirow{13}{*}{\scriptsize (2)} & \multirow{7}{*}{\shortstack{Trid.\\Reduct.}}  & \multirow{3}{*}{\trid}  & \num{\trid}{1} & $Q^TCQ = \ph{.}{T}$ & {\sc dsytrd} & LAPACK\\ 
			& & & \num{\trid}{2} & $TZ = Z\Lambda \rightarrow T,Z$ & {\sc dstemr} & LAPACK\\ 
			& & & \num{\trid}{3} & $Y := QZ$ & {\sc dormtr} & LAPACK\\\cline{3-7}
& & \multirow{4}{*}{\trit}  & \num{\trit}{1} & $Q_1^TCQ_1 = \ph{.}{W}$ & {\sc dsyrdb} & SBR \\ 
		        & & & \num{\trit}{2} & $Q_2^TWQ_2 = T$ & {\sc dsbrdt} & SBR\\ 
		        & & & \num{\trit}{3} & $TZ = Z\Lambda \rightarrow T,Z$ & {\sc dstemr} & LAPACK\\ 
		        & & & \num{\trit}{4} & $Y := Q_1Q_2Z$ & {\sc dormtr} & LAPACK \\\cline{2-7}
& \multirow{8}{*}{\shortstack{Krylov\\Subsp.}}  & \multirow{3}{*}{\krye}  & \num{\krye}{1}  & $z_{k+1} := Cw_k$ & {\sc dsymv} & BLAS\\
		        & & & \num{\krye}{2} & $z_{k+1}\rightarrow w_{k+1}$ & {\sc dsaupd} & ARPACK \\            
		        & & & \num{\krye}{3} & $T_m, V_m   \rightarrow \Lambda,Y$ & {\sc dseupd} & ARPACK \\ \cline{3-7}
& & \multirow{5}{*}{\kryi}  & \num{\kryi}{1} & $\bar{w}_{k} := U^{-1}\ph{;}{w_k}$ & {\sc dtrsv} & BLAS\\ 
                        & & & \num{\kryi}{2} & $\hat{w}_{k} := A\bar{w}_k$ & {\sc dsymv} & BLAS \\ 
                        & & & \num{\kryi}{3} & $z_{k+1} := U^{-T}\hat{w}_k$ & {\sc dtrsv} & BLAS \\
                        & & & \num{\kryi}{4} & $z_{k+1}\rightarrow w_{k+1}$ & {\sc dsaupd} & ARPACK \\ 
                        & & & \num{\kryi}{5} & $T_m, V_m   \rightarrow \Lambda,Y$ & {\sc dseupd} & ARPACK \\ \hline
{\scriptsize (3)}    & {--}       & {--} & {\num{bt}{1}} & {$\ph{.}{X} := U^{-1}Y$} & {{\sc dtrsm}} & {BLAS}\\ \hline
\multicolumn{7}{c}{}\\
\multicolumn{7}{c}{\scriptsize (1): Reduction to standard, {\sc gs}. (2): Standard Eigenvalue Problem. (3): Back-transform, {\sc bt}.}\\
\end{tabular} 
\end{center}
}
\caption{Routines from conventional libraries necessary to build the \geneig\ solvers   
  for multi-core processors.
         \label{tab:keys}}
\end{table}


\begin{table}
{\footnotesize 
\begin{center}
\begin{tabular}{|c||c|c|c|c||c|c|c|c|} 
\hline
 \multirow{2}{*}{Key} & \multicolumn{4}{c||}{Experiment 1 (MD), $s$=100} & \multicolumn{4}{c|}{Experiment 2 ({\sc DFT}), $s$=448} \\ \cline{2-9}
 & \trid & \trit & \krye & \kryi & \trid & \trit & \krye & \kryi \\ \cline{2-9}
\hline \hline
\num{gs}{1}    & ~6.60 & ~6.60 & ~6.60 & ~6.60 & ~ 36.42 & ~ 36.42 & ~ 36.42 & ~ 36.42 \\  
\num{gs}{2}    & 27.54 & 27.54 & 27.54 &   --   & 140.35 & 140.35 & 140.35 & --   \\ \hline  
\num{\trid}{1} & 67.39 &      --    &  --    &     --   & 342.01 & --   & --   & --  \\ 
\num{\trid}{2} & ~0.54 &     --    &  --    &     --   & ~~4.57 & --   & --   & --  \\ 
\num{\trid}{3} & ~0.86 &    --    &   --    &     --   & ~~7.81 & --   & --   & -- \\ \hline  
\num{\trit}{1} &      --    & 54.47 &   --    &    --   &     --     & 272.86 & --   & -- \\ 
\num{\trit}{2} &      --    & 93.16 &    --    &    --   &     --     & 375.67 & --   & -- \\ 
\num{\trit}{3} &     --    & ~0.54 &    --    &    --   &     --     &  ~~4.57 & --   & --  \\ 
\num{\trit}{4} &     --    & ~0.46 &    --    &    --   &     --     & ~~4.53 & --   & -- \\ \hline
\num{\krye}{1} &   --    &   --    & ~4.72 &    --   &     --     & --   & 200.65 & -- \\     
\num{\krye}{2} &   --    &   --    & ~0.53 &    --   &    --     & --   & 107.44 & -- \\ 
\num{\krye}{3} &   --    &   --    & ~0.18 &    --   &    --     & --   & ~13.38 & -- \\ \hline 
\num{\kryi}{1} &   --    &   --    &    --    & 13.92 &    --     & --   & --   & 645.93 \\  
\num{\kryi}{2} &   --    &   --    &    --    & ~4.72 &    --     & --   & --   & 214.07  \\  
\num{\kryi}{3} &   --    &   --    &    --    & 13.56  &    --    & --   & --   & 618.37 \\  
\num{\kryi}{4} &   --    &   --    &    --    & ~0.54   &    --    & --   & --   & 118.29 \\ 
\num{\kryi}{5} &   --    &   --    &    --    & ~0.18   &    --    & --   & --   & ~13.74 \\ \hline 
\num{bt}{1}    & ~0.31 & ~0.31 & ~0.31 & ~0.31 & ~~2.41 & ~~2.41 & ~~2.41 & ~~2.41 \\ \hline \hline 
{\bf Tot.}    & {\bf 103.24} & {\bf 183.08} & {\bf 39.88} & {\bf 39.83} & {\bf 533.57} & {\bf 836.81} & {\bf 500.65} & {\bf 1,649.23} \\ \hline
\end{tabular} 
\end{center}
}
\caption{Execution time (in seconds) of the \geneig\ solvers
  on multi-core processors.\label{tab:timekernelMD}}
\end{table}

\subsection{Experimental evaluation}

Table~\ref{tab:timekernelMD} reports the execution time of the four
eigensolvers \trid, \trit, \krye, and \kryi\ for the solution of both MD's and DFT's \geneig\
on the multi-core platform.
The solvers are implemented using routines 
from the conventional software libraries listed above,
and compute 100 ($\approx$1\%) and 448 ($\approx$2.6\%) eigenpairs
for the MD and DFT experiments, respectively. 
These values reflect the needs of the associated application.

In Experiment~1, the execution time for the two variants of the Krylov-subspace
approach is approximately the same.  The number of
ARPACK iterations that the Krylov-based variants require for this
particular eigenproblem, (288 for both \krye\ and \kryi,) basically
balance the higher cost per iteration of \kryi\ (13.92+13.56=27.48
seconds to compute the two triangular solves, in \num{\kryi}{1} and
\num{\kryi}{3}) with that of building explictly $C$ in \krye\ (27.54
seconds due to \num{gs}{2}).  The low performance of both
tridiagonal-reduction variants (\trid\ and \trit) can be credited to
the cost of the reduction to tridiagonal form. Theoretically, this
operation is not much more expensive than the transformation to
standard form (e.g., $4n^3/3$ flops for \trid\ versus $n^3/3$ flops
for the Cholesky factorization plus $n^3$ additional flops for the
construction of $C$). Nevertheless, the fact that half of the flops
performed in the reduction to tridiagonal form via a direct method
(variant \trid) are cast in terms of BLAS-2, explains the high execution time
of this operation on a multi-core processor.  Avoiding this type of
low-performance operations is precisely the purpose of variant
\trit\ but, at least for this experiment, the introduction of a large 
overhead in terms of additional number of flops (in the accumulation of 
$Q_1Q_2$ during the reduction from band to tridiagonal form)
destroys the benefits of using BLAS-3.  
Finally, it is worth mentioning that the execution time of
the tridiagonal eigensolver (operations \num{\trid}{2} and
\num{\trit}{2}) is negligible, validating the choice of MR$^3$ for this step.

The situation varies in Experiment~2. Now \krye\ is the fastest
variant, followed closely by the tridiagonal-reduction \trid. The
reason lies in the number of ARPACK iterations that the Krylov-based
variants requires for this eigenproblem (now quite high,
4,034 for \krye\ and 4,261 for \kryi), which increases considerably the
overall cost of the iterative stage, especially for \kryi.  Variant
\trit\ is not competitive, mainly due to the cost of the accumulation
of orthogonal transformations during the reduction of the band matrix
to tridiagonal form (operation \trit2). For this particular problem
and value of $s$, the MR$^3$ algorithm applied to the tridiagonal
eigenproblem adds only a minor cost to the execution time.

\begin{table}
{\small 
\begin{center}
\renewcommand{\arraystretch}{1.5}
\begin{tabular}{|c||c|c|c|c|} 
\hline
 \multirow{2}{*}{~} & \multicolumn{4}{c|}{Experiment 1 (MD), $s=$100} \\ \cline{2-5}
 & \trid & \trit & \krye & \kryi \\ \cline{2-5}
\hline \hline
$ \| I -X^T \bar{B} X\|_{F}  \over \ph{}{\|}\bar{B}\|_{F} $ &   6.68E-21   &   6.56E-21   &   5.58E-21   &   6.73E-21 \\
$ \| \bar{A} X - \bar{B} X \Lambda\|_{F}  \over \ph{{\tiny :}}{\max}( \|\bar{A}\|_{F}, \|\bar{B}\|_{F}) $   &   1.03E-16   &  1.03E-16  &   1.05E-16   &   3.80E-16 \\
\hline
\end{tabular} 
\end{center}
}

{\small 
\begin{center}
\renewcommand{\arraystretch}{1.5}
\begin{tabular}{|c||c|c|c|c|} 
\hline
 \multirow{2}{*}{~} & \multicolumn{4}{c|}{Experiment 2 ({\sc DFT}), $s=$448} \\ \cline{2-5}
 & \trid & \trit & \krye & \kryi \\ \cline{2-5}
\hline \hline
$ \| I -X^T \bar{B} X\|_{F}  \over \ph{}{\|}\bar{B}\|_{F} $ &  1.15E-15   &   2.29E-14   &   1.35E-15   &   1.43E-15  \\
$ \| \bar{A} X - \bar{B} X \Lambda\|_{F}  \over \ph{{\tiny :}}{\max}( \|\bar{A}\|_{F}, \|\bar{B}\|_{F}) $ &   9.80E-16   &  1.93E-15  &   6.45E-16   &   1.93E-14   \\
\hline
\end{tabular} 
\end{center}
}
\caption{Accuracy of the \geneig\ solvers built from conventional libraries.
         \label{tab:accuracyCPU}}
\end{table}

Table~\ref{tab:accuracyCPU} shows the accuracy of the solutions (in
terms of relative residual and orthogonality) obtained with the four
eigensolvers. In Experiment~1, our algorithms are applied to the
inverse eigenpair $(\bar{A},\bar{B})=(B,A)$, as computing its $s$
largest eigenpairs yields faster convergence in this case; in 
Experiment~2, $(\bar{A},\bar{B})=(A,B)$.  The results show that the
accuracy of \trid\ and \krye\ are comparable but there exists a slight
degradation of variant \kryi, which may be due to the operation with
the upper triangular factor $U$ at each iteration.

\begin{figure}
\begin{center}
\begin{tabular}{cc}
\begin{minipage}[c]{0.46\textwidth}
\includegraphics[width=\textwidth]{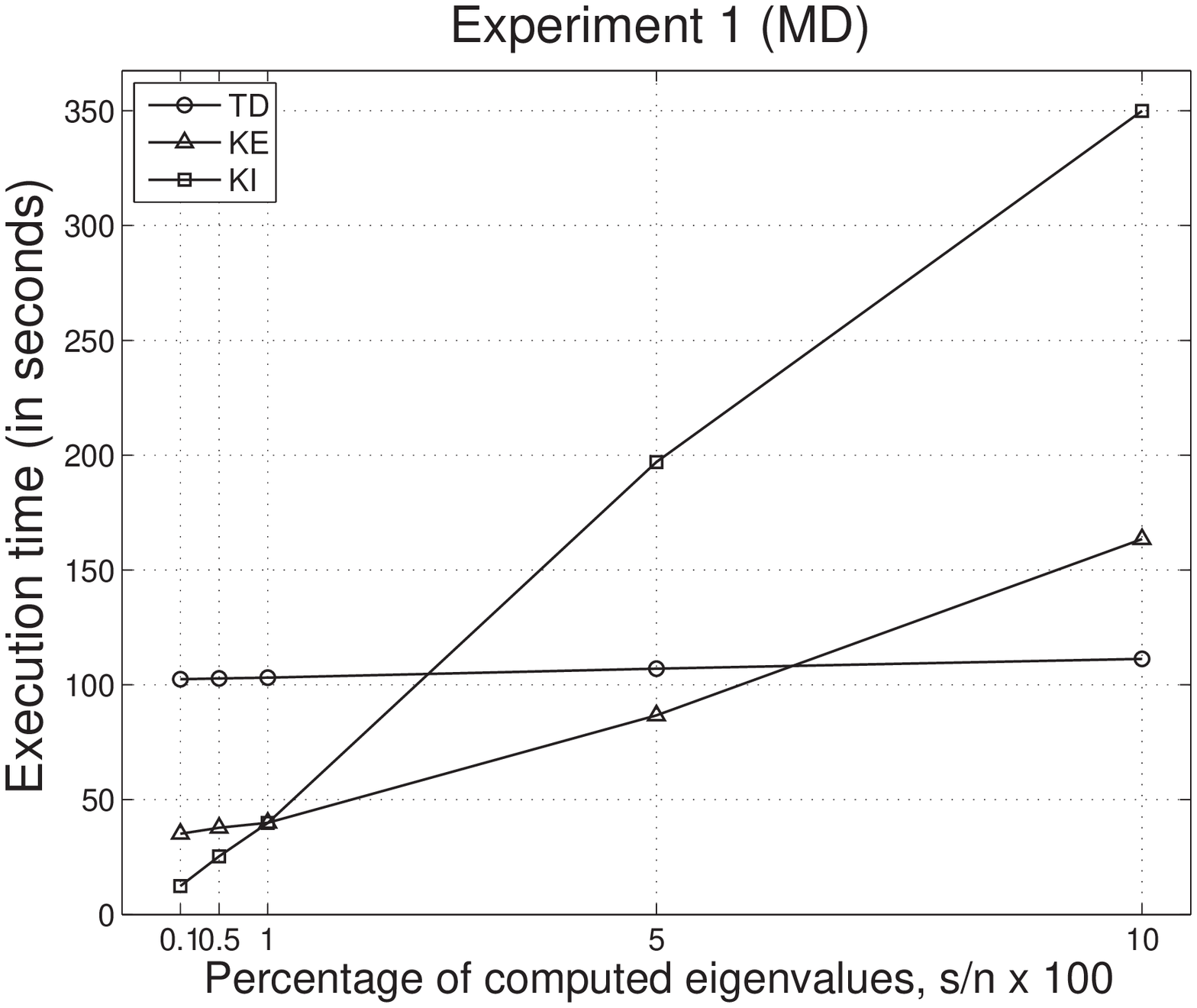}
\end{minipage}
\begin{minipage}[c]{0.49\textwidth}
\includegraphics[width=\textwidth]{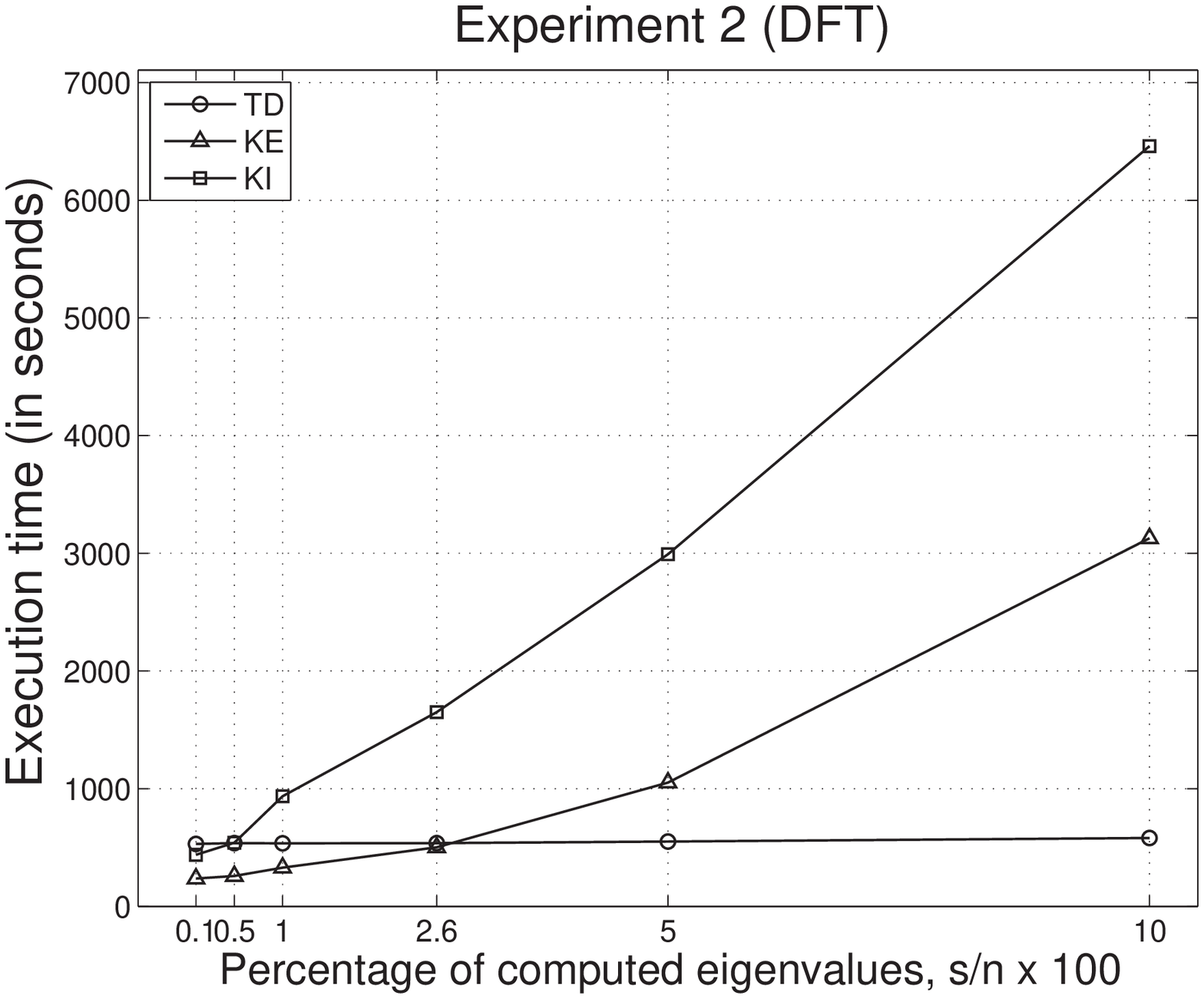}
\end{minipage}
\end{tabular}
\end{center}
\caption{\label{fig:timevss}Execution time of the \geneig\ solvers on multi-core 
  processors for different values of the number of computed eigenpairs $s$.}
\end{figure}

To close our evaluation of the implementations based on conventional
libraries, Figure~\ref{fig:timevss} reports the execution times of
variants \trid, \krye, and \kryi\ for different values of $s$. 
(Variant \trit\ is not included because the
previous experiments clearly demonstrated that it was not
competitive.)  The results show a rapid increase in the execution
time of the variants based on the Krylov-subspace as $s$ grows, due to a
significant increase in the number of steps that these iterative
procedures require as well as the increment in the costs associated with re-orthogonalization 
and restart, which respectively grow quadratically and linearly with $m$ (with $m > 2s$). 
This is particularly penalising for variant \kryi,
due to its higher cost per iteration. The small increase in the
execution time of variant \trid, on the other hand, is mostly due to
the back-transform.


\section{Libraries for Multi-threaded Architectures\label{sec:modern_libraries}}

\subsection{Task-parallel libraries for multi-core processors}

With the emergence of multi-core processors,
and especially with the increase in the number of processing elements 
in these architectures, exploiting task-level
parallelism has been recently reported as a successful path to improve the performance of both dense linear
algebra operations~\cite{CPE:CPE1463,Buttari200938,SuperMatrix:TOMS} and 
sparse linear system solvers~\cite{Aliaga2011183};
moreover, projects like Cilk, SMPSs, and StarPU have demonstrated the assets of leveraging task-based parallelism in
more general computations.
Modern dense linear algebra libraries that adhere to the task-parallel approach
include PLASMA and {\tt libflame}+SuperMatrix (hereafter \supermatrix).
Unfortunately, the current releases of PLASMA (2.4.2) and \supermatrix\ (5.0) provide only a reduced number of kernels,
and for the generalized eigenvalue problem, they only
cover the initial reduction to \stdeig. 
Concretely, \supermatrix\ provides routines {\sc FLA\_Chol} and {\sc FLA\_Sygst} for operations \num{gs}{1} and \num{gs}{2},
while PLASMA implements only routine {\sc PLASMA\_dpotrf} for the first operation.
The performance of these kernels are compared with those of LAPACK/BLAS in Table~\ref{tab:timekernelCPU}.
The results there show that the use of these task-parallel libraries especially benefits those variants 
which explictly construct $C$ (\trid, \trit\ and \krye). In particular, if we consider the effect of the reduction
of the execution time of \num{gs}{2} using \supermatrix, \krye\ becomes clearly faster than \kryi\ in 
Experiment~1. In the other experiment, the situation does not vary, as the faster solvers were
\trid\ and \krye, and they equally benefit from any improvement to the construction of $C$.

\begin{table}[hb]
{\footnotesize 
\begin{center}
\begin{tabular}{|c||c|c|c||c|c|c|} 
\hline
 \multirow{2}{*}{Key} & \multicolumn{3}{c||}{Example 1 (MD), $s=$100} & \multicolumn{3}{c|}{Example 2 ({\sc DFT}), $s=$448} \\ \cline{2-7}
 & LAPACK/BLAS & \supermatrix & PLASMA & LAPACK/BLAS & \supermatrix & PLASMA \\ \cline{2-7}
\hline \hline
\num{gs}{1}    & ~6.60  & ~5.63  & 5.13 & 36.42  & 25.19 & 27.97 \\ 
\num{gs}{2}    & 27.54  & 14.18  & --   & 140.35 & 83.34 & -- \\ \hline  
\end{tabular} 
\end{center}
}
\caption{Execution time (in seconds) of the task-parallel eigensolvers 
         on multi-core processors.
         \label{tab:timekernelCPU}}
\end{table}

\subsection{Kernels for GPUs}

The introduction of GPUs with unified architecture and programming style~\cite{CUDA} posed
quite a revolution for the scientific and high-performance community. 
Linear algebra was not an exception, 
and individual efforts~\cite{CPE:CPE1472,Volkov2008} were soon
followed by projects (e.g., \supermatrix, MAGMA, CULA~\cite{culaweb})
conducted to improve and extend the limited functionality (and sometimes performance) 
of the implementation of the BLAS from NVIDIA (CUBLAS).

\begin{table}
{\small 
\begin{center}
\begin{tabular}{|c|c|c|cc|c|c|} 
\hline
Stage & Appr.  &  Var. &      & Operation & Routine(s) & Library\\ \hline \hline
\multirow{5}{*}{(1)} & \multirow{5}{*}{--}       & \multirow{5}{*}{--} & \multirow{2}{*}{\num{gs}{1}}  & \multirow{2}{*}{$\ph{.}{B}=U^TU \rightarrow U$} & {\scriptsize {\sc magma\_dpotrf} or } & {\scriptsize {\sc MAGMA} or }       \\ 
                     &                           &                     &              &                                & {\scriptsize {\sc fla\_chol}}      & {\scriptsize {\sc \supermatrix}} \\ 
                     &                           &                     & \multirow{3}{*}{\num{gs}{3}}  & \multirow{3}{*}{$C:=U^{-T}AU^{-1}$}             & {\sc fla\_sygst}                   & \supermatrix                     \\ 
                     &                           &                     &              &                                & {\scriptsize {\sc cublasDtrsm} or }   & {\scriptsize {\sc CUBLAS} or }      \\ 
                     &                           &                     &              &                                & {\scriptsize {\sc magma\_dtrsm}}   & {\scriptsize {\sc MAGMA}}        \\ \hline
\multirow{13}{*}{(2)} & \multirow{7}{*}{\shortstack{Trid.\\Reduct.}}  & \multirow{3}{*}{\trid}  & \num{\trid}{1} & $Q^TCQ = \ph{.}{T}$ & {\sc magma\_dsytrd} & MAGMA\\ 
			& & & \num{\trid}{2} & $TZ = Z\Lambda \rightarrow T,Z$ & -- & -- \\ 
			& & & \num{\trid}{3} & $Y := QZ$ & -- & -- \\\cline{3-7}
& & \multirow{4}{*}{\trit}  & \num{\trit}{1} & $Q_1^TCQ_1 = \ph{.}{W}$ & {\sc GPU\_dsyrdb} & SBRG \\ 
		        & & & \num{\trit}{2} & $Q_2^TWQ_2 = T$ & {\sc GPU\_dsbrdt} & SBRG\\ 
		        & & & \num{\trit}{3} & $TZ = Z\Lambda \rightarrow T,Z$ & -- & -- \\ 
		        & & & \num{\trit}{4} & $Y := Q_1Q_2Z$ & -- & -- \\\cline{2-7}
& \multirow{12}{*}{\shortstack{Krylov\\Subsp.}}  & \multirow{4}{*}{\krye}  & \multirow{2}{*}{\num{\krye}{1}}  & \multirow{2}{*}{$z_{k+1} := Cw_k$}  & {\scriptsize {\sc cublasDsymv} or } & {\scriptsize {\sc CUBLAS} or }\\
&                                                 &                         &                 &                   & {\scriptsize {\sc magma\_dsymv}} & {\scriptsize {\sc MAGMA}}\\
		        & & & \num{\krye}{2} & $z_{k+1}\rightarrow w_{k+1}$ & -- & -- \\ 
                        & & & \num{\krye}{3} & $T_m, V_m   \rightarrow \Lambda,Y$ & --           & --     \\ \cline{3-7}
& & \multirow{8}{*}{\kryi}  & \multirow{2}{*}{\num{\kryi}{1}} & \multirow{2}{*}{$\bar{w}_{k} := U^{-1}\ph{:}{w_k}$} & {\scriptsize {\sc cublasDtrsv} or } & {\scriptsize {\sc CUBLAS} or }\\ 
& &                         &                &                                    & {\scriptsize {\sc magma\_dtrsv}} & {\scriptsize {\sc MAGMA}}\\ 
                        & & & \multirow{2}{*}{\num{\kryi}{2}} & \multirow{2}{*}{$\hat{w}_{k} := A\bar{w}_k$} & {\scriptsize {\sc cublasDsymv} or } & {\scriptsize {\sc CUBLAS}} or  \\ 
                        & & &                &                             & {\scriptsize {\sc magma\_dsymv}} & {\scriptsize {\sc MAGMA}} \\ 
                        & & & \multirow{2}{*}{\num{\kryi}{3}} & \multirow{2}{*}{$z_{k+1} := U^{-T}\hat{w}_k$} & {\scriptsize {\sc cublasDtrsv} or }  & {\scriptsize {\sc CUBLAS} or } \\
                        & & &                &                              & {\scriptsize {\sc magma\_dtrsv}} & {\scriptsize {\sc MAGMA}} \\
                        & & & \num{\kryi}{4} & $z_{k+1}\rightarrow w_{k+1}$ & -- & -- \\ 
                        & & & \num{\kryi}{5} & $T_m, V_m   \rightarrow \Lambda,Y$ & --           & --     \\ \hline
\multirow{2}{*}{(3)}     & \multirow{2}{*}{--} & \multirow{2}{*}{--}       & \multirow{2}{*}{\num{bt}{1}} & \multirow{2}{*}{$\ph{.}{X} := U^{-1}Y$} & {\scriptsize {\sc cublasDtrsm} or }  & {\scriptsize {\sc CUBLAS} or } \\
        &    &          &             &                        & {\scriptsize {\sc magma\_dtrsm}} & {\scriptsize {\sc MAGMA}}\\ \hline
\multicolumn{7}{c}{}\\
\multicolumn{7}{c}{\scriptsize (1): Reduction to standard, {\sc gs}. (2): Standard Eigenvalue Problem. (3): Back-transform, {\sc bt}.}\\
\end{tabular}
\end{center}
}
\caption{Routines from modern libraries necessary to build 
  the \geneig\ solvers
  for multi-threaded architectures.\label{tab:software2}}
\end{table}

\begin{table}[tb]
{\small 
\begin{center}
\begin{tabular}{|c||c|c|c|c||c|c|c|c|} 
\hline
 \multirow{2}{*}{Key} & \multicolumn{4}{c||}{Experiment 1 (MD), $s=$100} & \multicolumn{4}{c|}{Experiment 2 ({\sc DFT}), $s=$448} \\ \cline{2-9}
 & \trid & \trit & \krye & \kryi & \trid & \trit & \krye & \kryi \\ \cline{2-9}
\hline \hline
\num{gs}{1}    & ~1.52 & ~1.52 & 1.52 & 1.52 & ~~7.12 & ~~7.12 & ~ ~7.12 & ~ ~7.12 \\ 
\num{gs}{2}    & ~7.38 & ~7.38 & 7.38 & -- & ~44.17 & ~44.17 & ~44.17 & --   \\ \hline  
\num{\trid}{1} & 59.08 & -- & --  & --  & 297.84  & -- & -- & --  \\ 
\num{\trid}{2} & {\bf ~0.54} & -- & -- & -- & {\bf ~~4.57} & -- & -- & --  \\ 
\num{\trid}{3} & {\bf ~0.86} & -- & --  & --  & {\bf ~~7.81}  & -- & -- & -- \\ \hline  
\num{\trit}{1} & -- & 31.60 & --  & --  & -- & 152.37   & -- & -- \\ 
\num{\trit}{2} & -- & 47.70 & --  & --  & -- & ~92.18   & -- & -- \\ 
\num{\trit}{3} & -- & {\bf ~0.54} & --  & --  & --  & {\bf ~~4.57} & -- & --  \\ 
\num{\trit}{4} & -- & {\bf ~0.46} & --  & --  & -- & {\bf ~~4.53} & -- & -- \\ \hline
\num{\krye}{1} & -- & -- & ~1.79 & --  & -- & -- & ~75.31 & -- \\ 
\num{\krye}{2} & -- & -- & {\bf ~0.46} & --  & -- & -- & {\bf 123.97} & -- \\ 
\num{\krye}{3} &   --    &   --    & {\bf ~0.18} &    --   &    --     & --   & {\bf ~13.17} & -- \\ \hline 
\num{\kryi}{1} & -- & -- & --  & 10.64  & --  & -- & -- & 296.73 \\ 
\num{\kryi}{2} & -- & -- & --  & ~1.79  & --  & -- & -- & {\bf 210.77}  \\ 
\num{\kryi}{3} & -- & -- & --  & 11.06  & --  & -- & -- & 310.47 \\
\num{\kryi}{4} & --  & --  & --  &  {\bf ~0.54}  & --  & -- & -- & {\bf 121.89} \\ 
\num{\kryi}{5} &   --    &   --    &    --    & {\bf ~0.18}   &    --    & --   & --   & {\bf ~13.17} \\ \hline 
\num{bt}{1}    & ~0.05 & ~0.05 & ~0.05  & ~0.05  & ~~0.84 & ~~0.84 & ~~0.84 & ~~0.84 \\ \hline
{\bf Tot.} & {\bf 69.43} & {\bf 89.25} & {\bf 11.38} & {\bf 25.78} & {\bf 362.35} & {\bf 305.76} & {\bf 264.58} & {\bf 970.12} \\ \hline
\end{tabular} 
\end{center}
}
\caption{Execution time (in seconds) of the conventional+modern \geneig\ solvers 
         on multi-threaded architectures.
         The numbers in boldface are
         obtained using the LAPACK in place of the missing GPU routines. 
         \label{tab:timekernelGPU}}
\end{table}

As of today, the development of dense linear algebra libraries 
for GPUs is still immature, but certain kernels exist and have demonstrated 
performance worth of being investigated. Table~\ref{tab:software2}
contains a list of GPU kernels related to the
solution of \geneig s. 
The MAGMA and CUBLAS libraries provide
routines for the Cholesky factorization, the tridiagonalization, as
well as several Level 2 and 3 BLAS operations. The reduction from
\geneig\ to \stdeig\ is implemented in \supermatrix, while routines for 
the two-stage tridiagonalization (SBRG) were developed as part of 
previous work~\cite{CPE:CPE1680,PDP:Davor2011}.

\subsection{Experimental evaluation of prototype libraries}

Table~\ref{tab:timekernelGPU} reports the execution time of the four eigensolvers employing
the kernels specified in Table~\ref{tab:software2}. Those operations for which no GPU kernel 
was available were computed on the CPU and the corresponding timings are marked in bold face in the table.
In this case, the time required to transfer the data between the main memory and the hardware 
accelerator's memory space is included in the result. (Because of the transfer cost, the timings for 
the operations performed on the CPU are not the same as those reported  
in Table~\ref{tab:timekernelMD}).

Whenever a GPU kernel was provided by more than a library
(e.g., routines {\sc fla\_sygst} from \supermatrix, {\sc cublasDtrsm} from CUBLAS or 
{\sc magma\_dtrsm} from MAGMA) we selected the one included in MAGMA.
The timings using kernels from \supermatrix\ for operation \num{gs}{1} were slightly worse than
those obtained with MAGMA for both Experiments~1 and~2. On the other hand, \supermatrix\ outperformed
the kernel in MAGMA for \num{gs}{2} in Experiment~2 but was inferior in Experiment~1.
CUBLAS offered similar or worse performance in all these cases.

The first observation to make is the remarkable difference between the
execution time of variant \krye\ when the GPU is employed to
accelerate operation \num{gs}{2} in Experiment~1: from 27.54 to only
7.28 seconds (a speed-up of 3.73) using, in this case, two calls to
the triangular system solve from MAGMA. This is complemented by
the lowering of the timing for operation \num{gs}{1} using the
Cholesky factorization from MAGMA; for this operation, GPUs attain an even 
higher speed-up, 4.34, but on a less dominant stage. 
Combined, the two stages lead to an overall 3.5$\times$ acceleration factor 
of variant \krye, which is now the best method for this experiment. 
While other variants also have to compute the same operations, 
the acceleration reported by the GPU is blurred by the minor cost 
of \num{gs}{} compared with other operations. 
Experiment~2 also benefits from the use of the GPU during
the initial transformation from \geneig\ to \stdeig.
However, for the large experiment and variant \kryi, we
cannot use the matrix-vector products in this platform.  The matrices
involved in this experiment are too large to keep two $n \times n$
arrays into the GPU memory, one for the triangular factor $U$ and one
for $A$.  


While the GPU promises important gains when applied to perform 
CPU-bound computations (with intensive data parallelism), in some cases 
the results are somewhat disappointing. This is the case, for example, 
of the reduction to tridiagonal form in variant \trid, which applies the 
routine {\sc dsytrd} (from MAGMA library) that shows much slower speed-up 
on the GPU than expected.  On the other hand, our GPU 
implementation of variant \trit\ attains much better performance than \trid, 
although it is still slower than the Krylov-based approach \krye.
Besides, the general evaluation is that in some
cases the GPU routines in these libraries are not directly applicable
as, e.g., happens when the data matrices are too large to fit into
the device memory, which in general is much smaller than the main
memory. This requires a certain knowledge of the numerical operation,
to transform it into a sort of out-of-core routine. While such a restructuring is
easy for some operations like the triangular system solve with
multiple right-hand sides, dealing with others like the reduction to
band form turns out to be quite a complex task~\cite{PDP:Davor2011}.

\begin{table}[tb]
{\small 
\begin{center}
\renewcommand{\arraystretch}{1.5}
\begin{tabular}{|c||c|c|c|c|} 
\hline
 \multirow{2}{*}{~} & \multicolumn{4}{c|}{Experiment 1 (MD), $s=$100} \\ \cline{2-5}
 & \trid & \trit & \krye & \kryi \\ \cline{2-5}
\hline \hline
$ \| I -X^T \bar{B} X\|_{F}  \over \|\bar{B}\|_{F} $ &   4.02E-20   &   4.01E-20   &   4.04E-20   &   4.03E-20   \\
$ \| \bar{A} X - \bar{B} X \Lambda\|_{F}  \over \max( \|\bar{A}\|_{F}, \|\bar{B}\|_{F}) $   &   5.41E-16   &  5.42E-16  &   5.41E-16   &   5.72E-16 \\
\hline
\end{tabular} 
\end{center}
}

{\small 
\begin{center}
\renewcommand{\arraystretch}{1.5}
\begin{tabular}{|c||c|c|c|c|} 
\hline
 \multirow{2}{*}{~} & \multicolumn{4}{c|}{Experiment 2 ({\sc DFT}), $s=$448} \\ \cline{2-5}
 & \trid & \trit & \krye & \kryi \\ \cline{2-5}
\hline \hline
$ \| I -X^T \bar{B} X\|_{F}  \over \|\bar{B}\|_{F} $ & 1.61E-14   &   3.68E-14   &   1.42E-15   &   1.38E-15  \\
$ \| \bar{A} X - \bar{B} X \Lambda\|_{F}  \over \max( \|\bar{A}\|_{F}, \|\bar{B}\|_{F}) $   &   5.41E-15   &  1.56E-15  &   7.46E-16   &   5.33E-14   \\
\hline
\end{tabular} 
\end{center}
}
\caption{Accuracy of the  conventional+modern \geneig\ solvers. 
         \label{tab:accuracyGPU}}
\end{table}

In Table~\ref{tab:accuracyGPU} we report the accuracy of
the \geneig\ eigensolvers built on top of the conventional+modern libraries. 
In Experiment~1, all methods yield similar results while, in Experiment~2, 
the iterative solvers present slightly better accuracies.
On the other hand, 
in general there are little qualitative differences between the results obtained with conventional 
and the conventional+modern libraries.

\begin{figure}
\begin{center}
\begin{tabular}{cc}
\begin{minipage}[c]{0.46\textwidth}
\includegraphics[width=\textwidth]{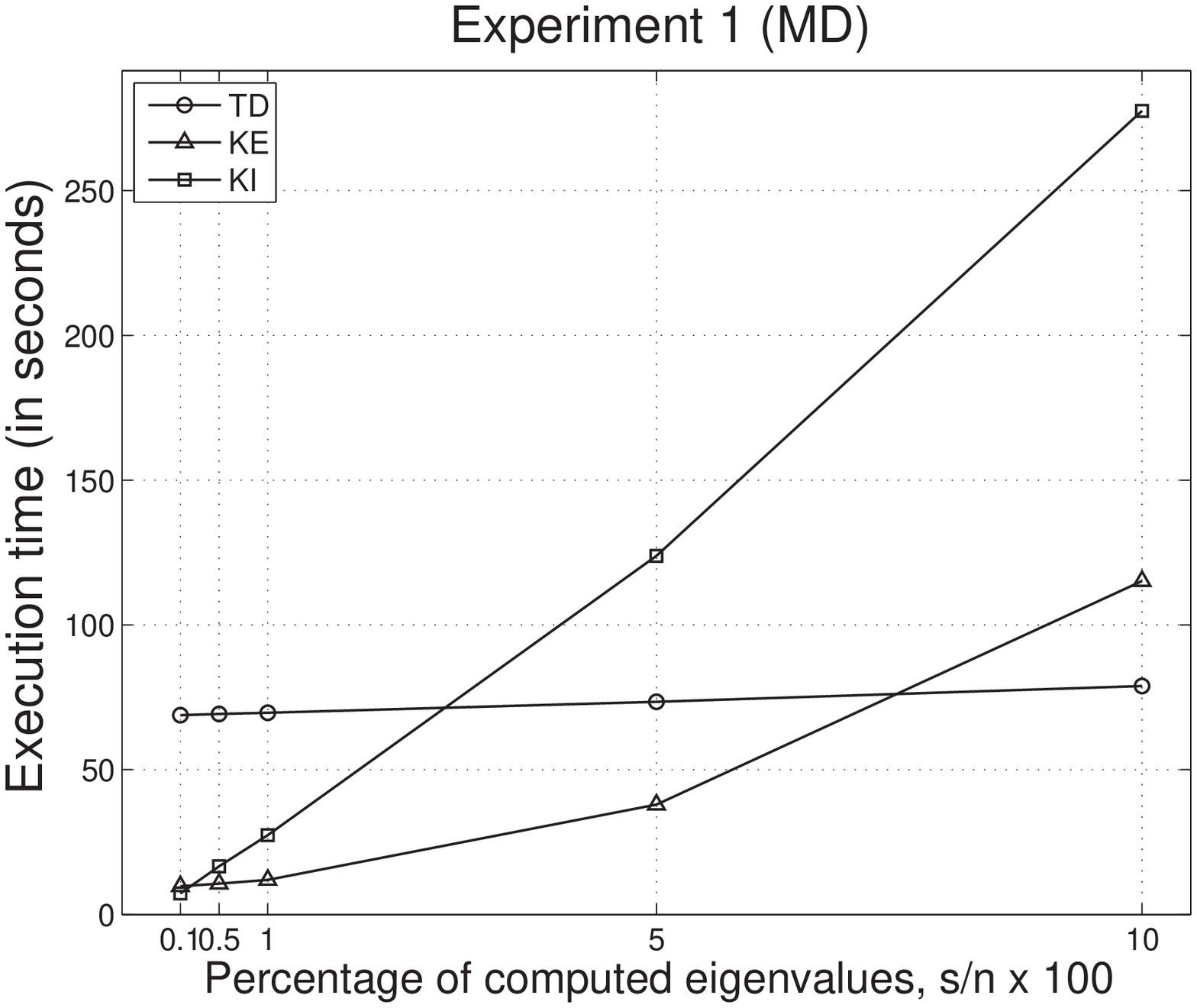}
\end{minipage}
\begin{minipage}[c]{0.49\textwidth}
\includegraphics[width=\textwidth]{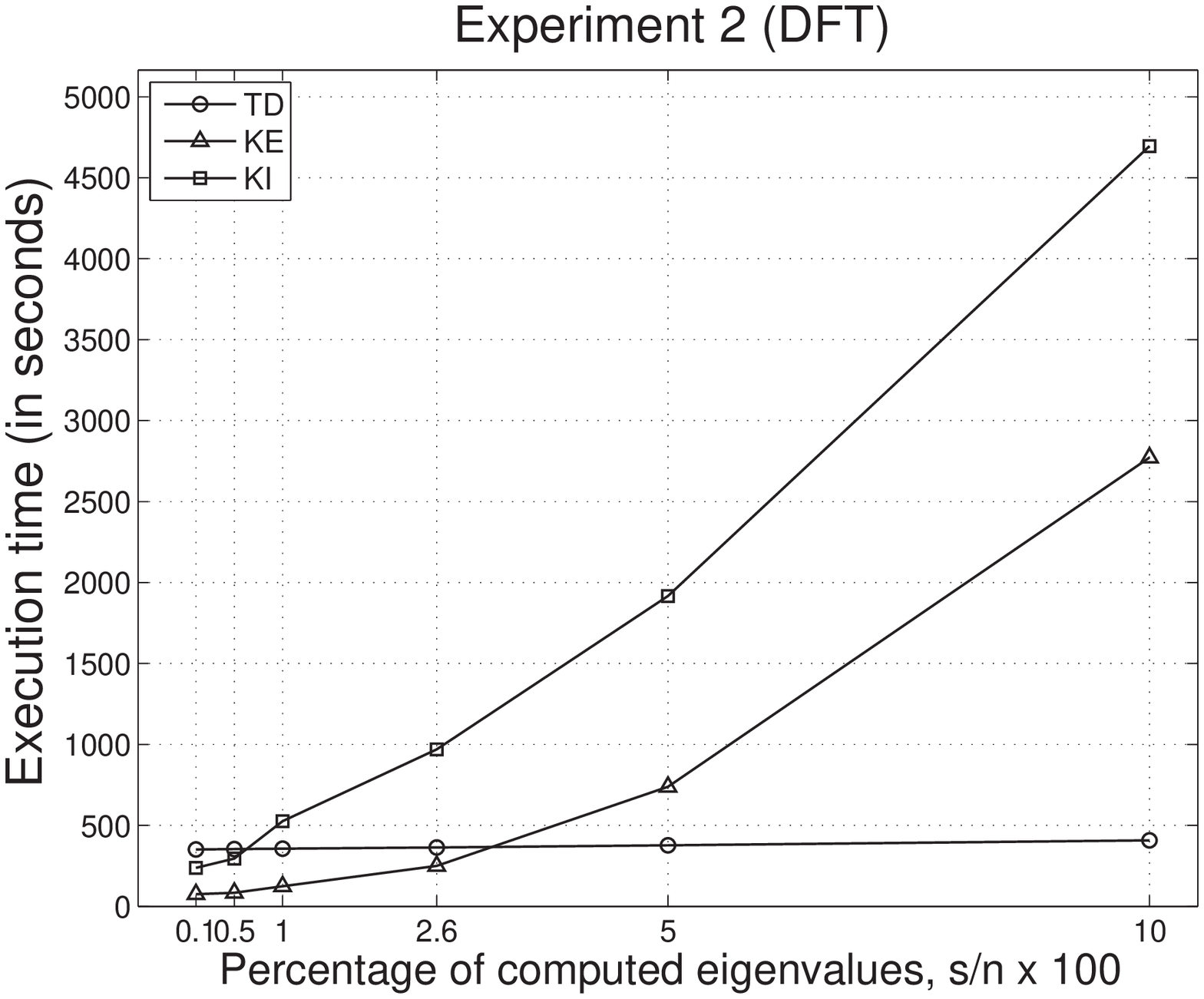}
\end{minipage}
\end{tabular}
\end{center}
\caption{\label{fig:timevss_GPU}Execution time of the conventional+modern \geneig\ solvers 
  for different values of the number of computed eigenpairs $s$.}
\end{figure}

Figure~\ref{fig:timevss_GPU} re-evaluates the performance of variants
\trid, \krye, and \kryi\ as a function of $s$, now leveraging 
the implementation of these solvers using the kernels 
in the conventional+modern libraries. The results exhibit
a rapid increase in the execution
time of the variants based on the Krylov-subspace with the dimension of~$s$, due to the
growth in the number of iterative steps, especially for \kryi.

\section{Conclusions\label{sec:conclusions}}

We presented a performance study for the solution of generalized
eigenproblems on multi-threaded architectures.  The focus was on two
different approaches: the reduction to tridiagonal form ---either
directly or in successive steps---, and the iterative solution through
a Krylov method. In both cases, we first built the eigensolvers on top
of conventional numerical libraries (BLAS, LAPACK, SBR, and ARPACK),
and then compared with implementations that make use of modern
multi-threaded libraries ({\tt libflame}, PLASMA, MAGMA, and CUBLAS)
as well as a few GPU kernels that we developed ourselves. As testbeds,
we chose matrices arising in large-scale molecular dynamics and
density functional theory; in both applications, only a portion of the
lower part of the spectrum is of interest. The results are representative
of the benefits that one should expect from GPUs and multi-threaded
libraries; moreover, they indicate that in realistic applications,  
when only 3--5\% of the spectrum is required, 
the Krylov-subspace solver is to be preferred.


\bibliographystyle{plain}
\bibliography{biblio,enrique}

\end{document}